\documentclass[12pt]{article}
\usepackage{amssymb} \usepackage{amsfonts}
\usepackage{graphicx}
\begin{document}
\title{\bf{Landau parameters for isospin asymmetric nuclear matter based on a
relativistic model of composite and finite extension nucleons}
\thanks{\it{This work was partially supported by the CONICET, Argentina.}}}
\author{R. M. Aguirre and A. L. De Paoli.\\
Departamento de F\'{\i}sica, Fac. de Ciencias Exactas,\\
Universidad Nacional de La Plata.\\
C. C. 67 (1900) La Plata, Argentina.}
\maketitle
\begin{abstract}

We study the properties of cold asymmetric nuclear matter at high
density, applying the quark meson coupling model with excluded
volume corrections in the framework of the Landau theory of
relativistic Fermi liquids. We discuss the role of the finite
spatial extension of composite baryons on dynamical and
statistical properties such as the Landau parameters, the
compressibility, and the symmetry energy. We have also calculated
the low lying collective eigenfrequencies arising from the
collisionless quasiparticle transport equation, considering both
unstable and stable modes. An overall analysis of the excluded
volume correlations on the collective properties is performed.\\


\end{abstract}
\newpage

\renewcommand{\theequation}{\arabic{section}.\arabic{equation}}

\section{Introduction}

The study of the nuclear medium composed of different fractions of
protons and neutrons has been developed for long time and it has
concentrated a renewed interest in the last years. The equation of
state of isospin asymmetric nuclear matter is a subject of
particularly intense research \cite{VARIOS0, VARIOS1,
VARIOS2,SEEXP, AVAN}. The possible applications range from the
structure of radiative nuclei, the dynamics of rare isotopes and
the cooling process of neutron stars. A key role in many of these
calculations is played by the density dependence of the symmetry
energy \cite{VARIOS2}, which may be extracted from recent isospin
diffusion data in heavy ion collisions  experiences
\cite{VARIOS1}.\\
The standard theoretical calculations have used as the relevant
degrees of freedom protons and neutrons, moving
non-relativistically through potentials representing the averaged
instantaneous interaction among nucleons. Furthermore, since the
development of the Quantum Hadro-Dynamics \cite{SW}, it has became
a common practice to use relativistic hadronic fields in these
kind of calculations. In such a case the nuclear interaction is
mediated by mesons of scalar or vector isospin character. The role
of the scalar isovector meson ($a_0$ (980)) has been emphasized in
recent investigations \cite{VARIOS4}.

A further step has been given by models incorporating the quark
structure of hadrons. They may be reduced to mean field hadronic
pictures whose parameters, such as masses and vertices, hide the
quark dynamics. These effective models provide both a connection
between the hadronic phenomenology and the fundamental theory of
the strong interactions, besides a sound description of hadronic
matter and atomic nuclei.

Among the dual quark-hadron theoretical frameworks, those based on
the bag models include explicitly the quark confinement volume.
Specifically, the Quark Meson Coupling (QMC) \cite{ST} states the
dynamical evolution of this confining region, which depends on the
global properties of the nuclear medium as well as on the
configuration of the hadronic fields. However, this feature is
usually missed \cite{ST} in passing to a pure hadronic context, as
baryons and
mesons are regarded as point-like particles in this limit.\\
The relevance of the finite extension of nucleons in the
evaluation of some nuclear statistical properties has been
stressed long time ago \cite{KAPUSTA}. In order to fulfill this
assertion, several corrections have been introduced in the
hadronic interactions, mainly invoking a Van der Waals-like
normalization \cite{Waa1,Cley,QMCCC,OUR}. The motivation for such
a procedure is to parameterize in compact form the strong
baryon-baryon repulsion at very short distances. In the high
density realm of nuclear matter the mean free path of nucleons is
comparable to the quark confining size, hence excluded
volume effects become relevant. \\
The correction due to the finite extension of nucleons have been
introduced into the hadronic sector of the bag models \cite{QMCCC,
OUR}, it was found that it is necessary to describe properly high
density hadronic matter without violating the model assumptions
\cite{OUR}.

Another theoretical scheme suitable to describe the nuclear
collective phenomena is the Landau theory of Fermi liquids.
Although it was originally stated in a non-relativistic fashion,
see for instance \cite{BAYMPETH}, it was subsequently extended to
deal with the relativistic fields formalism \cite{MATSUI,CAILLON}.
The Landau parameters are useful to evaluate thermodynamical
properties, the stability conditions in phase transitions, the
nuclear matter collective excitations which couple, for instance,
to the weak interaction governing the neutrino emission of URCA
processes in neutron stars.

In this work the behaviour of isospin asymmetric nuclear matter is
studied in the framework of the QMC model with excluded volume
corrections in the hadronic sector. Special attention is paid to
the Landau parameters and the collective modes. In the next
section we give a resume of the QMC model and we describe the
correlations generated by the Van der Waals-like normalization. In
section 3 we define the Landau parameters and use them in section
4 to derive the isoscalar and isovector collective modes. Results
and discussions are presented in section 5, and finally the
conclusions are drawn in section 6.

\section{The Quark Meson Coupling Model} \setcounter{equation}{0}

QMC is an effective model of the quark structure of hadrons, it is
inspired in the MIT bag model, so that in its starting point the
confinement mechanism has being accomplished and chiral symmetry
has been broken. In order to describe the dynamics of the emerging
hadrons, usually the model is projected into a picture of
point-like baryons interacting through virtual mesons \cite{ST},
the same
 ones which couple quarks inside the confinement region. Since we are primarily
 concerned with isospin asymmetric nuclear matter, we only consider $u$ and $d$ flavor
 of quarks coupled by the isovector $\rho$ and $\delta$ ($a_0$ (980)) mesons, besides
 the commonly used iso-scalar $\sigma$ and $\omega$ ones.\\
Within the QMC model baryons are represented as non-overlapping
spherical bags containing three valence quarks; the bag radius
changes dynamically with the fields configuration.\\
The mean field approximation (MFA) is the natural scheme of
solution, which replaces the meson fields by its classical
expectation values. These mean values determine the nucleon
fields, which in turn become the source of the meson
ones.\\
In the MFA the Dirac equation for a quark of flavor $q, \; (q=u,
d)$, of current mass $m_q$ and $I_q$ third isospin component, is
given by \cite{ST}

\begin{equation}
( i \gamma^{\mu} \partial_{\mu} - g_{\omega}^q \gamma^0 \omega - g_{\rho}^q I_q\; \gamma^0 \rho - {m_q}^\ast) \Psi^q = 0, \label{QMCEQ}
\end{equation}
here the notation ${m_q}^{\ast}=m_q - g_{\sigma}^q \sigma-I_q
g_\delta^q \delta$ is used.\\
For a spherically symmetric bag of radius $R_b$ corresponding to a baryon of class $b$,
 the normalized quark wave function
$\Psi^q_b(r,t)$ is given by

\begin{equation}
\Psi^q_b(r,t)={\cal N}_b^{-1/2} \frac{e^{-i{\varepsilon}_{q b} t}}
{\sqrt{4\pi}} \left( \begin{array}{c}
j_0 (x_{qb} \, r/R_b) \\
i \beta_{q b} {\vec{\sigma}}.{\hat{r}} j_1 (x_{q b} \, r/R_b)
\end{array} \right) \chi ^q,
\end{equation}
where $\chi ^q $ is the quark spinor and

\begin{equation}
\varepsilon_{q b} = \frac{\Omega_{q b}}{\!R_b} + g_{\omega}^q
\;\omega + g_{\rho}^q I_q\; \rho,
\end{equation}

\begin{equation}
{\cal N }_b={R_b}^3\;[2 \Omega_{q b} (\Omega_{q b} - 1) + R_b
{m_q}^\ast ]\; \frac{ j_0^2 (x_{q b}) }{x_{q b}^2},
\end{equation}

\begin{equation}
\beta_{q b} ={\left[ \frac{\Omega_{q b} - R_b {m_q}^\ast
}{\Omega_{q b} + R_b {m_q}^\ast } \right]}^{1/2},
\end{equation}
with $\Omega_{q b} =[x_{q b}^2 +{(R_b {m_q}^\ast)}^2    ]^{1/2}$.
The eigenvalue $x_{q b}$ is the lowest solution of the equation

\begin{equation}
j_0 (x_{q b}) = \beta_{qb} \; j_1 (x_{q b}), \label{BOUNDARY}
\end{equation}
which arises from the boundary condition at the bag surface.\\ In
this model the ground state bag energy is identified with the
baryonic mass $M_b$,

\begin{equation}
M_b=\frac{\sum_q n_q^b \Omega_{q b} - z_{0 b}}{R_b} +
\frac{4}{3} \pi B_0   {R_b}^3, \label{BAGMASS}
\end{equation}
where $n_q^b$ is the number of quarks of flavor $q$ inside the
bag. The bag constant $B_0$ is numerically adjusted to get definite
values for the proton bag radius, and the zero-point motion parameters $z_{0 b}$
are fixed to reproduce the baryon spectrum at zero density.\\
The dispersion relation for the b-baryon is
\begin{equation}
k_0^b=\sqrt{{M_b}^2+(\mathbf{k}-\mathbf{\Sigma}_b)^2} \pm
g_\omega \omega_0 \pm g_\rho I_b \rho_0^3, \label{PARTENER}
\end{equation}
for particle $(+)$ and antiparticle $(-)$ solutions. Within the
MFA at zero temperature only the particle solutions contribute. In
this expression we have assumed that the strength of the couplings
does not depend on the quark flavor. We have also introduced the
baryonic isospin projection $I_b=\sum_q n_q^b I_q$, and the vector
nucleon selfenergy $\mathbf{\Sigma}_b=g_\omega
\mathbf{\omega}+g_\rho \mathbf{\rho} I_b$.

For homogeneous infinite static matter the spatial dependence of all the meson fields can be neglected, so that its equations of
motion reduce in the MFA to
\begin{eqnarray}
\sigma&=&=-\frac{1}{{m_\sigma}^2} \sum_b
\left( \frac{\partial M_b}{\partial\sigma}\right)_{R,\delta} n_s^b, \label{SIGMA} \\
\omega_\mu&=& \frac{1}{{m_\omega}^2}
\sum_b g_{\omega} j_\mu^{\;b} , \label{OMEGA}\\
\delta&=&-\frac{1}{{m_\delta}^2} \sum_b
\left(\frac{\partial M_b}{\partial \delta}\right)_{R,\sigma} n_s^b, \label{DELTA}\\
\rho_\mu&=&\frac{1}{{m_\rho}^2} \sum_b g_{\rho} I_b j_\mu^{\;b}.
\label{RHO}
\end{eqnarray}
Here $j_\mu^{\,b}=(n^b,\mathbf{j}^{\;b})$ stands for the mean
value of the nucleon current density of isospin $b$. In the
reference frame where the averaged momentum of matter is zero the
spatial components of the currents become null
$\mathbf{j}^{\;b}=0$, hence only the time-like
projections of the vector mesons are non-zero. \\
It must be noted that Eqs. (\ref{DELTA}) and (\ref{RHO}) refer
only to the third isospin component, since the remaining ones
become zero in the MFA.

From the relations (\ref{BAGMASS}) and (\ref{PARTENER}) it can be seen, as it was earlier
mentioned, that the nucleon mass and
energy spectrum depends on the assumed meson classical values. They are in turn, determined
 by the nucleon densities as given by the equations (\ref{SIGMA})-(\ref{RHO}).

The densities $n_s^b$ and $j_\mu^{\;b}$ are defined with respect
to the ground state of the hadronic matter, which at zero
temperature is composed of baryons filling the Fermi sea up to the
state with momentum $k_{Fb}$
\begin{eqnarray}
n_s^b &=& \;\frac{\vartheta} {(2\pi)^3} \sum_{spin} \int
\frac{d^3k\;M_b} {
\sqrt{{M_b}^2+(\mathbf{k}-\mathbf{\Sigma}_b)^2}}
\;\theta(k_{Fb}-|\mathbf{k}|),
\label{SCALDENS} \\
j_\mu^{\;b} &=& \; \frac {\vartheta} {(2 \pi)^3} \sum_{spin}\int
\frac{d^3k\;(k-\Sigma)_\mu} {
\sqrt{{M_b}^2+(\mathbf{k}-\mathbf{\Sigma}_b)^2}} \;
\theta(k_{Fb}-|\mathbf{k}|). \label{VECDENS}
\end{eqnarray}
In Eqs.  (\ref{SCALDENS}) and (\ref{VECDENS} ) the factor
$\vartheta$ is included for future use and it takes the value
$\vartheta = 1$ for point-like baryons.

The total energy density $\mathcal{E}$ and pressure $P_0$ of
hadronic matter for point-like baryons is evaluated as
\begin{eqnarray}
\mathcal{E}&=&\frac {1}{2} m_\sigma^2\; \sigma^2 +\frac {1}{2}
m_\omega^2\; \omega^2 +\frac {1}{2} m_\rho^2\; \rho^2+\frac {1}{2}
m_\delta^2\; \delta^2 \nonumber
\\ &+& \frac{\vartheta}{{\pi}^2} \sum_b
\int_0^{k_{Fb}} dk k^2 \sqrt{{M_b}^2+k^2}, \label{ENERGY} \\
P_0&=&\sum_b \mu_0^b n^b - \mathcal{E}, \label{PRESSURE}
\end{eqnarray}
\noindent
 where $\mu_0^b=k_0^b (k_{Fb})$, see Eq. (\ref{PARTENER}), is the chemical
 potential for point-like baryons.

In the QMC the radius $R_b$ is a variable dynamically adjusted to
reach the equilibrium of the bag in the dense hadronic medium. We use the equilibrium condition proposed in ref. \cite{OUR}, which can be obtained by minimizing the energy density $\mathcal{E}$
with respect to $R_b$

\begin{equation}
- \frac {1} {4 \pi R_b^2} {\left( \frac {\partial{M_b}}
{\partial R_b} \right)}_{\sigma, x_{q b}} =
 \frac{1}{3{\pi}^2 \, \xi} \sum_c \int_0^{k_{Fc}}
\frac{dk k^4} {\sqrt{{M_c}^2+k^2}}, \label{QMCc}
\end{equation}
where $\xi=1$. This result reflects the balance of the internal
pressure of the bag with the baryonic contribution to the total
external pressure, represented by the left and right sides of
Eq.(\ref{QMCc}), respectively.\\
The factor $\xi$ will be redefined below where excluded volume
effects will be considered.

The QMC model heavily relies on the assumption of non-overlapping
bags, using this criterion an upper density limit around three
times the saturation density of symmetric nuclear matter has been found \cite{SR}. In order
 to properly take into account this
severe restriction, a simplified model was introduced in
\cite{OUR}, which  describe baryons as extended objects. Since
finite size baryons are assumed non-overlapping, their motion must
be restricted to the available space $V'$ defined as \cite{Waa1}
\begin{equation}
V'=V-\sum_b N^b v_b,
\end{equation}
with $N^b$ the total number of baryons of class $b$ inside the
volume $V$, and $v_b$ the effective volume per baryon of this
class. The last mentioned quantity is proportional to the actual
baryon volume, i.e. for spherical volumes of radius $R_b$
\begin{equation}
v_b=\alpha \frac{4 \pi}{3} {R_b}^3,
\end{equation}
where $\alpha$ is a real number ranging from $4$, in the low
density limit, to $3\sqrt{2}/{\pi}$, which corresponds to the
maximum density allowed for non overlapping spheres, in a face
centered cubic arrange. Since we wish to study the high density
regime of homogeneous isotropic matter, we shall adopt
$\alpha= 3\sqrt{2}/{\pi}$ in all our calculations.

Consequently it was assumed that the nucleon fields can be
normalized replacing
 $V'$ for $V$, which in turn implies a normalization of the nucleon densities
  (\ref{SCALDENS}), (\ref{VECDENS}) and of the nucleon contribution to the
  energy (\ref{ENERGY}). The final result may be obtained by taking
   $\vartheta=1-\sum_b n^b  v_b$ within these equations and
   $\xi= {n^b_s}/{(\alpha \, n^b)}$ in Eq. (\ref{QMCc}) \cite{OUR}.\\

Since $\vartheta$ introduces an explicit dependence upon the
baryonic densities, the chemical potentials get an extra term,
i.e.
\begin{eqnarray}
\mu^b&=&{ \left( \frac{\partial \mathcal{E}}{\partial n_b} \right)
}_ {{ n_{{b'}_{{}_{{}_{{}_{\!\!\!\!\!\!\!\!\!\!\!b' \neq b}}}}}}}
= \mu_0^b + \Delta \mu^b, \label{PQ} \\ \Delta \mu^b &=&\frac{v_b
}{3{\pi}^2} \sum_c \int_0^{k_{Fc}} \frac{dk k^4}
{\sqrt{{M_c}^2+k^2}}. \label{POTQUIM2}
\end{eqnarray}

Correspondingly, the total pressure acquires an additional term
$\Delta P$ as compared to the pressure of point-like baryons $P_0$ in Eq. (\ref{PRESSURE})
\begin{eqnarray}
P_H=P_0 + \Delta P = P_0 + \sum\limits_{b}  n^b \Delta \mu^b.
\label{TRUEPRESS}
\end{eqnarray}

\section{Landau parameters and equation of state}
\setcounter{equation}{0}

The Fermi liquid theory of Landau assumes that the low-lying
excitations of a physical system admits a representation in terms
of quasi-particles and, circumstantially, collective modes. If the
quasi-particle states can be identified by a composed label $B=(b,
\beta)$, where we have singled out the first place for the isospin
projection and $\beta$ collects discrete spin and momentum
indices, then the occupation number of such a level is denoted by
$f_B$. The conserved baryonic number can be expressed in terms of
a summation over such distribution functions: $n=\sum_B f_B$. The
same statement is valid for every extensive conserved quantity
such as the energy, which can be written as $\sum_B\,f_B
\varepsilon_B$ plus current-current interactions, where
$\varepsilon_B$ is the nucleon single particle spectrum. For
infinite nuclear matter the summation over the discrete momentum
indices must be replaced
 by an integration over the continuous spectrum with measure $d^3k/(2 \pi^3)$.\\
The formulae (\ref{SCALDENS}) and (\ref{VECDENS}) for the nucleon
densities must be rewritten in order to fit this form. Firstly the
replacement $p_i = k_i\,\vartheta^{1/3}$ is made,
 secondly we identify $\sum_{spin} \int
d^3p\;\theta(p_{Fb}-|\mathbf{p}|)/(2\pi)^3
\rightarrow \sum_\beta f_{b\,\beta}$, with $p_{Fb} = k_{Fb}\, \vartheta^{1/3}$. Hence, Eqs.
(\ref{SCALDENS}), (\ref{VECDENS}), and (\ref{ENERGY}) can be
rewritten as
\begin{eqnarray}
n_s^b&=&\sum_\beta
\frac{f_{b\,\beta}\,M_b}{\sqrt{M_b^2+
(\vartheta^{-1/3}\mathbf{p}_\beta-\mathbf{\Sigma}_b)^2}},
\\
j_\mu^{\;b}&=&\sum_\beta \frac{f_{b\,\beta}\,(\vartheta^{-1/3} p_\beta -\Sigma_b)_\mu}
{\sqrt{M_b^2+(\vartheta^{-1/3}\mathbf{p}_\beta-\mathbf{\Sigma}_b)^2}}, \\
\mathcal{E}&=&\sum_B
f_{b\,\beta}\left(\sqrt{M_b^2+(\vartheta^{-1/3}\mathbf{p}_\beta-\mathbf{\Sigma}_b)^2}+
g_\omega \, \omega_0+ I_b g_\rho \rho_0 \right)\nonumber \\
&&+\frac{1}{2}\left(m_\sigma^2 \, \sigma^2+m_\delta^2 \,
\delta^2-m_\omega^2 \, \omega_\mu \omega^\mu-m_\rho^2 \, \rho_\mu
\rho^\mu \right),
\end{eqnarray}
where the momentum independence of the nucleon masses,
characteristic of the Hartree approximation, has been emphasized.\\
In the last formula $\mathcal{E}$ must be considered as a function
of the distribution functions $f_B$, the bag radii $R_b$ and the
meson fields $\phi=\sigma,\, \omega, \, \delta, \, \rho$, hence
first variation of the energy density yields
\[
\delta \mathcal{E}(f,R,\phi)=  \sum_b \frac{\partial
\mathcal{E}}{\partial R_b} \delta R_b + \sum_\phi \frac{\partial
\mathcal{E}}{\partial \phi} \delta \phi +\sum_B\frac{\partial
\mathcal{E}}{\partial f_B}\, \delta f_B.\nonumber
\]
The first term is null because of the equilibrium condition for
the bag in the nuclear medium, Eq. (\ref{QMCc}), the second one is likewise zero due to the
 meson field equations
(\ref{SIGMA})-(\ref{RHO}). Therefore, the quasiparticle energy
spectrum according to the Fermi liquid prescription is
\begin{eqnarray}
\varepsilon_B&=&\frac{\partial \mathcal{E}}{\partial
f_B}=\sqrt{M_b^2+(\vartheta^{-1/3}\mathbf{p}_\beta-\mathbf{\Sigma}_b)^2}+
g_\omega \omega_0+g_\rho \rho_0 I_b\nonumber  \\
&& +\frac{v_b}{3}\sum_{B'} f_{B'} \frac{\vartheta^{-4/3}\mathbf{p}_{\beta'}
\cdot\left(\vartheta^{-1/3}\mathbf{p}_{\beta'}-\mathbf{\Sigma}_{b'}\right)}{\sqrt{M_{b'}^2+(\vartheta^{-1/3}\mathbf{p}_{\beta'}-\mathbf{\Sigma}_{b'})^2}}. \label{QUASPECTRA}
\end{eqnarray}
If Eq. (\ref{QUASPECTRA}) is evaluated at the Fermi surface, {\it
i.e.} $|\mathbf{p}_\beta|=p_\beta=p_{Fb}$, in the limit of
isotropic matter and restoring the continuum spectrum, it yields
the chemical potential given by Eqs. (\ref{PQ}), (\ref{POTQUIM2}),
in support of the thermodynamical consistency of our approach.

In order to evaluate the Landau's parameters of the nuclear interaction, a second variation
 must be performed, i.e. $F_{B'\,B}=\partial^2 \mathcal{E}/\partial
f_{B'}\,\partial f_B=F_{B\,B'}$. Since all of the bag radii and
the meson fields depend ultimately on the distribution functions
$f_B$, Eqs. (\ref{SIGMA})-(\ref{RHO}) and (\ref{QMCc})
must be differentiated in order to obtain a closed expression. \\
The Landau's parameters are defined as the Fourier coefficients of
an expansion in terms of the Legendre polynomials
\[
F^l_{BB'}=(l+1/2)\int_{-1}^1 dx\,P_l(x)\,\frac{\partial^2
\mathcal{E}(x)}{\partial f_{B'}\,\partial f_B},\nonumber
\]
where the integrand must be evaluated on the Fermi surface at the
end of the calculations.\\
It is found that the second order derivative in the expression
above, depends linearly on the variable $x=\mathbf{p}_{\beta'}
\cdot \mathbf{p}_\beta/(p_{\beta'}\,p_\beta)$, hence all the
parameters of order greater than one are null, whereas for
$l=0,\,1$ we have
\begin{eqnarray}
F_{ab}^0&=&C_\omega+I_a I_b C_\rho+\sum_{c=n,p} \frac{n_c
v_c}{3\vartheta R_c} \left(3\frac{k_{Fa}^2}{E_a}+v_a H^{(3)}+v_a
\vartheta H^{(1)}
H_c^{(2)}\frac{M_c}{n_{sc}}\right)X_{cb}\nonumber
\\
&&+\frac{1}{3\vartheta}\left(\frac{k_{Fa}^2}{E_a} v_b+
\frac{k_{Fb}^2}{E_b} v_a
\right)+\frac{v_a}{R_a}\left(1-\frac{n_a}{n_{sa}}
\frac{M_a}{E_a}\right)H^{(1)}X_{ab} \nonumber \\
&&+v_a v_b \frac{H^{(3)}}{9 \vartheta}+\sum_{q=u,d}\left(n_a^q
Q_{qa} \frac{M_a}{E_a}-\frac{v_a}{3}\sum_{c=n,p}
n_c^q Q_{qc} M_c H_c^{(2)}\right)Y_{qb},\label{LPARAM0}\\
 F_{ab}^1&=&- \frac{k_a k_b}{E_a
 E_b}\;\frac{C_\omega+I_a I_b C_\rho+2 C_\omega C_\rho (1+I_a I_b)n_{\overline{a}}/E_{\overline{a}}}
 {1+(C_\omega+C_\rho)\sum_{c=p,n}\frac{n_c}{E_c}+4 C_\omega C_\rho \frac{n_1 n_2}{E_1
 E_2}}\label{LPARAM1}.
\end{eqnarray}
Here the definitions $C_\phi=(g_\phi/m_\phi)^2$ for
$\phi=\omega,\, \rho$ and $E_a=\sqrt{M_a^2+k_{Fa}^2}$, have been used, together
with
\begin{eqnarray}
H_a^{(1)}=\int_0^{k_{Fa}} \frac{dk k^4}{\pi^2
\left(M_c^2+k^2\right)^{1/2}},&&H_a^{(3)}=\frac{k_{Fa}^5}{\pi^2
E_a}\nonumber \\
H_a^{(2)}=\int_0^{k_{Fa}} \frac{dk k^4}{\pi^2
\left(M_c^2+k^2\right)^{3/2}},&&
H^{(l)}=\sum_{c=p,n}H_c^{(l)},\,l=1,2,3 \nonumber \\
Q_{q\,a}=\frac{\Omega_{qa}+2{m_q}^{\ast}R_a\left(\Omega_{qa}-1\right)}
{{m_q}^{\ast}R_a+2\Omega_{qa} \left(\Omega_{qa}-1\right)}.&&
\nonumber
\end{eqnarray}

Furthermore, the variables $X_{ab}$ and $Y_{qb}$ are the
derivatives $X_{ab}=\partial R_a/\partial f_B$ and
$Y_{qb}=\partial {m_q}^{\ast}/\partial f_B$ evaluated in the isotropic
matter limit and taking all the momenta at the Fermi surface.
Actually, they are the solutions of an algebraic coupled system of equations
obtained by
differentiating Eqs. (\ref{QMCc}),
(\ref{SIGMA}), and (\ref{DELTA}) and making a linear combination
of the two last results.\\
For further development, we define symmetric adimensional Landau parameters
\begin{equation}
\mathcal{F}_{a\,b}^{l}=\mathcal{F}_{b\,a}^{l}=\sqrt{\Gamma_a\,\Gamma_b}\,
F_{a\,b}^{l} \label{ADL}
\end{equation}
where $\Gamma_b=(k_{Fb} E_b/\pi^2)\,(b=p,n)$ is the relativistic
quasiparticle density of states at the Fermi surface.\\
Eqs. (\ref{LPARAM0})and (\ref{LPARAM1}) coincide with the previous
results \cite{MATSUI,CAILLON}  for symmetric nuclear matter if the
limit $v_a \rightarrow 0$ is taken carefully.

The Landau parameters can be related with the nuclear
compressibility $\kappa$ and symmetry energy $E_s$.  For a given
nucleon density $n=n^n+n^p$ and asymmetry coefficient
$t=(n^n-n^p)/n$, it can be shown that

\begin{equation}
\kappa=9\,n\,{\left(\frac{\partial^2\mathcal{E}}{{\partial
n}^2}\right)}_t = \frac{9\,n}{4} \left[
\frac{{(1-t)}^2}{\Gamma_p}+\frac{{(1+t)}^2}{\Gamma_n}\right]
\left(\frac{1}{\vartheta}+\mathcal{F}_{t}^{0}\right) \label{KAPPA}
\end{equation}
and
\begin{equation}
\frac{1}{2\,n}\,{\left(\frac{\partial^2\mathcal{E}}{{\partial
t}^2}\right)}_n= \frac{n}{8} \left(
\frac{1}{\Gamma_p}+\frac{1}{\Gamma_n}\right)
\left(\frac{1}{\vartheta}+\mathcal{F}_{s}^{0}\right) \label{SYM}
\end{equation}
where the effective Landau parameters $\mathcal{F}_{t}^{0}$ and
$\mathcal{F}_{s}^{0}$ are, respectively
\begin{eqnarray}
\mathcal{F}_{t}^{0}&=& \frac{{(1-t)}^2\,\Gamma_n\,\mathcal{F}_{p\,p}^0+
2\,(1-t^2)\,\sqrt{\Gamma_p\,\Gamma_n}\,\mathcal{F}_{p\,n}^0+
{(1+t)}^2\,\Gamma_p\,\mathcal{F}_{n\,n}^0}{{(1-t)}^2\,\Gamma_n+{(1+t)}^2\,
\Gamma_p} \nonumber \\
\mathcal{F}_{s}^{0}&=&
\frac{\Gamma_n\,\mathcal{F}_{p\,p}^0-2\,\sqrt{\Gamma_p\,\Gamma_n}\,
\mathcal{F}_{p\,n}^0+
\Gamma_p\,\mathcal{F}_{n\,n}^0}{\Gamma_n+\Gamma_p}
\end{eqnarray}
The symmetry energy $E_s(n)$ is obtained from Eq.(\ref{SYM}) in the limit of
$t=0$. These expressions reduce to the usual ones in the symmetric case without
volume corrections ($\vartheta=1$), see for instance \cite{MATSUI,CAILLON}.

\section{Collective excitations} \setcounter{equation}{0}

Collective modes are associated to local density fluctuations that
propagate in the hadronic mean field. These fluctuations are the
effect of small perturbations of the occupation distribution $f_B$
around the nucleon Fermi level of the quasiparticles. At zero
temperature the unperturbed nucleon distributions are given by
\cite{BAYMPETH}
\begin{equation}
f_B= \frac{1}{V}\,\theta(\mu_b-\varepsilon_B)
\end{equation}
where $V$ is the volume of the system, and $\mu_b$ is the chemical
potential. Local density fluctuations add a small variation of the
occupation numbers around the $f_B$ equilibrium value, i.e.
\cite{BAYMPETH,MATSUI}
\begin{equation}
f_B(\mathbf{r},t)=f_B+\delta f_B(\mathbf{r},t)=f_B+
\delta_{p_\beta\,p_{Fb}}\,u_b\,e^{i\,(\mathbf{q}.\mathbf{r}-\omega\,t)}
\label{FLUC} \end{equation} and to first order the quasiparticle
energies $\varepsilon_B$ change accordingly
\begin{equation}
\delta \varepsilon_B(\mathbf{r},t)=\sum_{B'} F_{B\,B'}\,
\delta f_{B'}(\mathbf{r},t)
\end{equation}
The propagation of these perturbations at low temperature is
governed by the collisionless Landau's kinetic equation
\begin{equation}
\frac{\partial f_B}{\partial t} + \frac{\partial f_B}
{\partial \mathbf{r}}\,\frac{\partial \varepsilon_B}{\partial
{\mathbf{p}}_\beta}-
\frac{\partial f_B}{\partial {\mathbf{p}}_\beta}
\,\frac{\partial \varepsilon_B}{\partial \mathbf{r}} = 0 \label{KINET}
\end{equation}
Introducing Eq. (\ref{FLUC}) into Eq. (\ref{KINET}) and keeping
only linear contributions of the fluctuations, we obtain
\begin{equation}
\frac{\partial \delta f_B}{\partial t} + \frac{\partial \delta f_B}
{\partial \mathbf{r}}\,\frac{\partial \varepsilon_B}{\partial
{\mathbf{p}}_\beta}-
\frac{\partial f_B}{\partial {\mathbf{p}}_\beta}
\,\frac{\partial \delta \varepsilon_B}{\partial \mathbf{r}} = 0 \label{LKINET}
\end{equation}
which can be further reduced to
\begin{equation}
\left[\omega-\vartheta^{-2/3}\,\frac{({\mathbf{p}}_{Fb}\,.\mathbf{q}\,)}{E_b}\right]
u_b-\frac{p_{Fb}}{\pi^2}\, ({\mathbf{p}}_{Fb}\,.\mathbf{q}\,)\,
\sum_{a=p,n} F_{b\,a}\,u_a=0 \label{CKIN}.
\end{equation}
The sum should be done at the Fermi surface of protons and
neutrons. For isotropic matter the direction of propagation can be
arbitrarily choosen along the azymuthal axis, so that
$({\mathbf{p}}_{Fb}\,.\mathbf{q}\,)={p}_{Fb}\;q\;cos\chi_b$.
Expanding the amplitudes $u_b$ and the Landau parameters
$F_{b\,a}$ in terms of Legendre polynomials
\begin{eqnarray}
u_b&=&\sum_{l}\,u_b^l\,P_l\left(\mbox{cos}\,\chi_b\right) \nonumber \\
F_{b\,a}^l&=&(2l+1)\int \frac{d\Omega_a}{4\,\pi}\,
F_{b\,a}\,P_l\left (\mbox{cos}(\chi_a-\chi_b)\right)
\end{eqnarray}
and making use of the addition theorem for Legendre polynomials, we have
\begin{equation}
\sum_{a=p,n} F_{b\,a}\,u_a=\sum_{a=p,n}\,\sum_{l}\,
\frac{1}{(2l+1)} F_{b\,a}^l\,u_a^l\,P_l
\left(\mbox{cos}\,\chi_b\right)
\end{equation}
Replacing in Eq. (\ref{CKIN}) and writing all back in terms of the
Fermi momenta $k_{Fb}$, we arrive to the following system of
homogeneus linear equations
\begin{equation}
\frac{u_b^l}{(2l+1)} +\vartheta\,\sum_{a=p,n}\, \sqrt{
\frac{\Gamma_b}{\Gamma_a}}\,\sum_{l'}\, \frac{1}{(2l'+1)}
\mathcal{F}_{b\,a}^{l'}\,u_a^{l'}\,\Omega_{l\,l'}(s_b)=0
\label{ZERO} \end{equation} with
$s_b=\vartheta^{1/3}\,(\omega/q)/v_{Fb}$, where
$v_{Fb}=(k_{Fb}/E_b)$ is the relativistic quasiparticle Fermi
velocity. The  function $\Omega_{l\,l'}$ is given by
\begin{equation}
\Omega_{l\;l'}(s)=\Omega_{l'\,l}(s)=\frac{1}{2} \int_{-1}^{\;\;1}dy\,P_l(y)\,
\frac{y}{(y-s)}\,P_{l'}(y)
\end{equation}
which take the particular expressions \cite{BAYMPETH}
\begin{eqnarray}
\Omega_{0\,0}(s)&=&1+\frac{s}{2}\,\ln\left(\frac{s-1}{s+1}\right) \nonumber \\
\Omega_{l\,1}(s)&=&s\,\Omega_{l\,0}(s) + \frac{1}{3} \delta_{l\,1}
\label{OME} \end{eqnarray} In the case of unstable modes for which
$s$ becomes a purely imaginary quantity in the upper half complex
plane, {\it i.e.} $s=i\zeta \;(\zeta
> 0)$, we have $\Omega_{0\,0}(i\zeta)=1-\zeta\;\mbox{arctan} ( 1/\zeta
)$. \\
Using Eq. (\ref{OME}) and keeping in mind that only terms
with $l=0,1$ are non vanishing, we can solve Eq. (\ref{ZERO}) for
the amplitudes $u_p^0$ and $u_n^0$
\begin{eqnarray}
u_p^0&=&\frac{C_p}{3\,s_p}\,u_p^1 + \frac{\vartheta}{9\,s_p}
\sqrt{\frac{\Gamma_p}{\Gamma_n}} \mathcal{F}_{p\,n}^{1}\,u_n^1 \nonumber \\
u_n^0&=&\frac{C_n}{3\,s_n}\,u_n^1 + \frac{\vartheta}{9\,s_n}
\sqrt{\frac{\Gamma_n}{\Gamma_p}} \mathcal{F}_{n\,p}^{1}\,u_p^1
\end{eqnarray}
where $C_b=(1+\frac{\vartheta}{3}\,\mathcal{F}_{b\,b}^{1}\,),
(b=p,n)$. Replacing into the equations for $l=1$, we have
\begin{eqnarray}
A_{p\,p}\,u_p^1+\sqrt{\frac{\Gamma_p}{\Gamma_n}}\,A_{p\,n}\,u_n^1=0 \nonumber \\
\sqrt{\frac{\Gamma_n}{\Gamma_p}}\, A_{n\,p}\,u_p^1+A_{n\,n}\,u_n^1=0 \label{SYST}
\end{eqnarray}
where
\begin{eqnarray}
A_{p\,p}&=&C_p+\vartheta\,\Omega_{0\,0}(s_p)(\,C_p\,\mathcal{F}_{p\,p}^{0}+
\vartheta\,\frac{v_{Fn}}{3\,v_{Fp}}\,\mathcal{F}_{p\,n}^{0}\,
\mathcal{F}_{n\,p}^{1}+{s_p}^2\,\mathcal{F}_{p\,p}^{1}\,) \nonumber \\
A_{p\,n}&=&\vartheta\,[\, \frac{1}{3}\mathcal{F}_{p\,n}^{1}+
\Omega_{0\,0}(s_p)(\,\frac{\vartheta}{3}\mathcal{F}_{p\,p}^{0}
\mathcal{F}_{p\,n}^{1}+\frac{v_{Fn}}{v_{Fp}}C_n\mathcal{F}_{p\,n}^{0}+
{s_p}^2\mathcal{F}_{p\,n}^{1}\,)\,]
\end{eqnarray}
The remaining coeficients $A_{n\,n}$, $A_{n\,p}$ are obtained
through the replacement $p \leftrightarrow n$ in the previous
formulas. The nontrivial eigenmodes $u_b^l(m), (m=1,2,...)$ of the
linearized transport equation (\ref{ZERO}) are equivalent to the
non vanishing solutions of Eq.(\ref{SYST}). Therefore the
corresponding eigenvalues $s_b^{(m)}$ satisfy
\begin{equation}
A_{p\,p}\,A_{n\,n} - A_{p\,n}\,A_{n\,p} = 0 \label{EIGEN}
\end{equation}
For fixed nucleon density $n$ and isospin asymmetry $t$, the solutions
$s_b^{(m)}$ of Eq. (\ref{EIGEN}) determine the zero sound dispersion relation
${(\omega/q)}^{(m)}$ for the $m$-eigenmode.
Following \cite{BARAN} we shall use the sign of the relative amplitudes
$\varrho= u_p^0/u_n^0$ to determine the isoscalar ($\varrho>0$) or isovector
($\varrho<0$) character of each mode, for arbitrary isospin asymmetry.

\section{Results and discussion} \setcounter{equation}{0}

\begin{figure}[ht]
\centering
\includegraphics[width=0.8\textwidth]{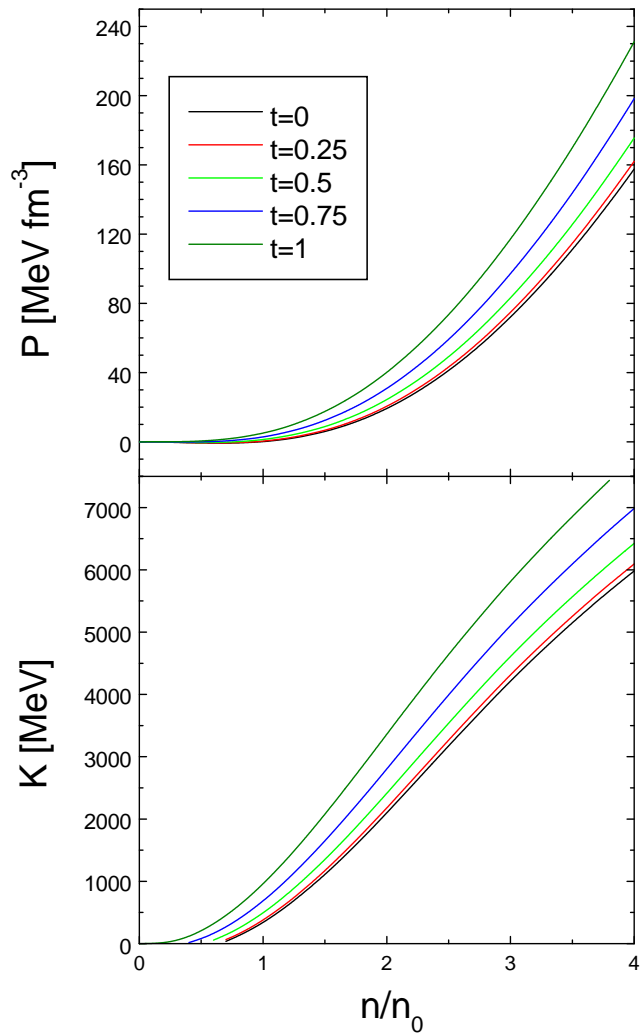}
\caption{\footnotesize{The pressure of cold asymmetric nuclear
matter versus density for the CC case (upper panel). The different
lines correspond to some typical values of the asymmetry parameter
$t$. In the lower panel the corresponding compressibility $\kappa$
is shown for the same set of $t$ values.}}\label{FIG1}
\end{figure}

\begin{figure}[ht]
\centering
\includegraphics[width=0.8\textwidth]{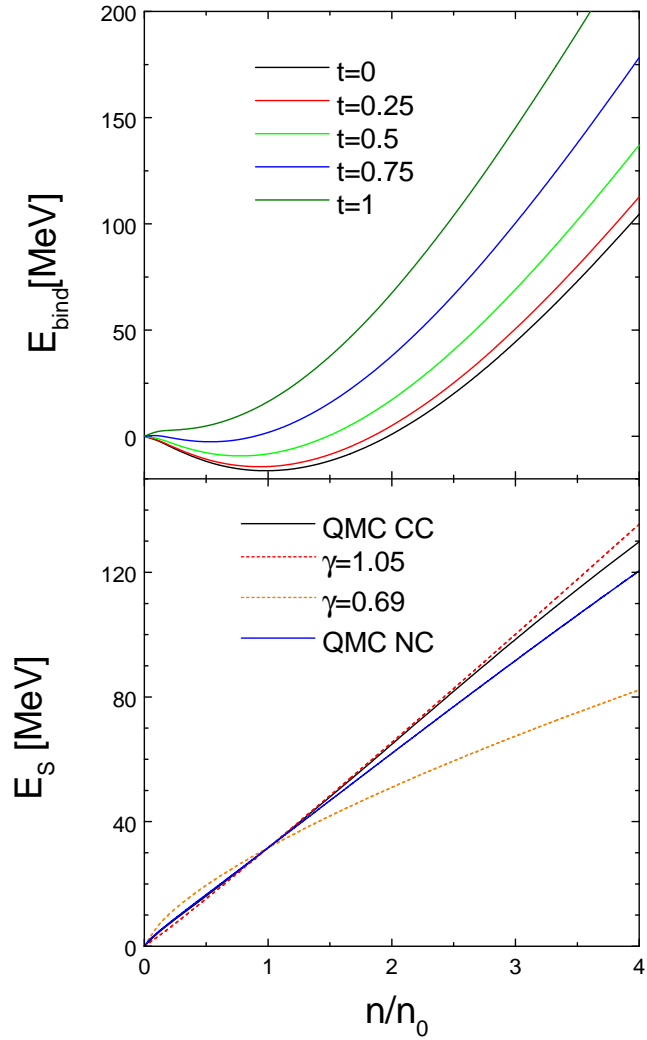}
\caption{\footnotesize{In the upper panel the binding energy of
cold asymmetric nuclear matter for the CC case and for some
typical values of the asymmetry parameter $t$ is drawn.
  In the lower panel we plot the asymmetry energy  for the CC and NC cases, together
  with the empirical expression  $E_s=31.6 \,(n/n_0)^\gamma$  ($\gamma=0.69, 1.05$)
   Ref.\cite{VARIOS1}.}}\label{FIG2}
\end{figure}

Within the present model the current quark masses have been chosen
as $m_u=m_d=5$ MeV, the bag parameter has been fixed as
$B_0^{1/4}=210.89$ MeV neglecting its eventual density dependence.
The parameter $z_{0b}$ was adjusted in order to reproduce at zero
density the empirical value of the nucleon mass and a nucleon bag
radius $R_b=0.6$ fm. Numerical values for the meson masses have
been taken as $m_\sigma=550$ MeV, $m_\omega=783$ MeV,
$m_\delta=984$ MeV and $m_\rho=770$ MeV.\\
Since mesons interact directly with quarks, the corresponding
meson-nucleon couplings are related to the quark-meson ones
$g_{\phi}^u=g_{\phi}^d$ ($\phi = \sigma, \omega, \delta, \rho $) in a simple way by assuming vector meson dominance, i.e.
\[
g_{\sigma}^{b}=3 g_{\sigma}^{u}, \;\;  g_{\omega}^{b} = 3
g_{\omega}^{u},\;\; g_{\delta}^{b} =g_{\delta}^{u},
\;\;g_{\rho}^{b} =g_{\rho}^{u}, \nonumber
\]
for $b=p,n$.  Their numerical values are obtained by reproducing
the symmetric nuclear matter saturation properties, i.e. baryonic
density, binding energy and symmetry energy
\begin{eqnarray}
& & n_0 = \;0.15 fm^{-3},  \nonumber \\ & & E_{bind}=
{(\mathcal{E}/n)}_0 -M c^2 = -16 \mbox{MeV}, \nonumber \\ & &
E_s(n_0)=\frac{1}{2\,n_0}\,{\left(\frac{\partial^2\mathcal{E}}{{\partial
t}^2} \right)}_{t=0} =\;31.6 \mbox{MeV},
\end{eqnarray}
where $M = 938.92 \mbox{MeV}/{c^2}$ is the average free nucleon
rest mass.\\
 The constraint of the symmetry energy alone is not
sufficient to determine unambiguously both $g_\rho$ and
$g_\delta$, instead it establishes a non-linear relation between
them. In order to restrain their possible variation, we have used
valuable phenomenological data coming from the analysis of isospin
diffusion in heavy ion
collisions \cite{VARIOS2}, as discussed below.\\
The values of the couplings are sensitive to either the inclusion
or not of the excluded volume corrections.
Both instances are considered in Table \ref{TABLEI}.\\
The cases with the excluded volume correction (CC) have been
compared with calculations without it (NC). It must be emphasized
that in the last case the values $\xi = \vartheta=1$, and
 $\Delta \mu^p=\Delta \mu^n=0$ must be used.\\
With this set of parameters the equation of state for asymmetric
nuclear matter with finite nucleon size has been evaluated. The
results are displayed  in Fig. \ref{FIG1} for the
 the pressure $P$ and the nuclear
compressibility $\kappa$ in terms of the baryonic number density
for several isospin asymmetries. For symmetric nuclear matter
($t=0$) at normal density we obtained $\kappa=340 MeV$, which is
higher than the usually assumed values in similar calculations.
However it should be stressed that recent mass measurements of the
pulsar PSR J0751+18007 yield $M=2.1\pm 0.2 M_\odot $ \cite{NICE},
which is hardly compatible with the usually adopted value $\kappa
\approx 250$ MeV. Indeed a stiffer equation of state is needed to
reach
this observational constraint. \\
In previous investigations \cite{OUR} the authors found that
finite baryonic size are responsible for a $12 \%$ increment in
the maximum mass of a neutron star, obtaining $M_{max}=1.89
M_\odot$. Therefore the strong baryonic repulsion previous to the
Quark-Gluon Plasma transition contributes significatively to
produce a stiffer equation of state, and consequently it would
give a phenomenological basis to understand the unexpectedly high
value measured for the PSR J0751+18007 mass.\\
 The compressibility becomes
negative at low densities leading to thermodynamical
instabilities,  but it grows with increasing $t$ at a fixed
density. The instabilities disappear for nuclear compositions
approaching pure neutron matter ($t=1$) as a
consequence of the repulsive character of the asymmetry energy.\\
The effect of increasing asymmetry on the binding energy
$E_{bind}$ can be appreciated in the upper panel of Fig.
\ref{FIG2}. In particular neutron matter remains unbound for all
densities. A comparison of the equation of state for the CC and NC
cases was given in \cite{OUR} for hadronic matter in
$\beta$-equilibrium, the  conclusions given there can be extended
to the case of nuclear matter at fixed isospin asymmetry, {\it i.
e.} density dependence for energy and pressure are stiffer
in the CC instance.\\
In the lower panel of the same figure the asymmetry energy $E_s$
is displayed, together with the curves corresponding to the
empirical expression $E_s=31.6 \,(n/n_0)^\gamma$ evaluated at the
limit values $\gamma=0.69$ and $\gamma=1.05$ obtained in
Ref.\cite{VARIOS1}. It can be seen that our result lies between
these curves, showing a significative agreement with the
$\gamma=1.05$ case in all the range $0.5 < n/n_0 < 2.5$. It must
be mentioned that within this model, only a narrow range of values
for the pair of couplings $g_\delta, g_\rho$ is able to fit the
reference value for $E_s(n_0)$ and to produce simultaneously a
curve entirely comprised between the phenomenological constraints.
For very low values of $g_\rho$ it is not possible to adjust the
symmetry energy at the normal density, increasing this coupling
yields a stiffer density dependence for $E_s(n)$, which quickly
goes beyond the curve
$\gamma=1.05$ of Fig.\ref{FIG2}.\\
Assuming a decomposition
\[
E_s(n)=E_s(n_0)+L(n/n_0-1)/3+K_s(n/n_0-1)^2/18,
\]
we have found $L\simeq 95.8$ MeV and $K_s\simeq 19$ MeV. The first
quantity agrees with the value $L=88\pm 25$ MeV found in
\cite{VARIOS0}, whereas we obtained for the combination
$K_{asy}=K_s-6\,L=-555\pm 5$ MeV in comparison with the suggested
value $K_{asy}=-500\pm 50$ MeV \cite{VARIOS1}. It must be pointed
out that the empirical value for $L$ is coherent with the
inequality $0.7<\gamma<1.1$, however the numerical value extracted
from experimental data for
$K_{asy}$ favors a stiffer symmetry energy with $1.26<\gamma<1.3$. \\
A comparison between NC and CC results yields a enhanced growth
for $E_s(n)$ in the last case, although differences become
appreciable for densities higher than $2 n_0$. For the NC case the
values $L=91.4 \pm 0.6$ MeV and $K_{asy}=-553 \pm 2$ MeV have been obtained.\\
In the limit of point-like baryons the formulae for the
compressibility and the symmetry energy reduce to
\begin{eqnarray}
\kappa_{t=0}=9 n\left[C_\omega+\frac{\pi^2}{2 k_F E}-\frac{C_\sigma^{eff}}
{1+C_\sigma^{eff}H^{(2)}} \left(\frac{M}{E}\right)^2\right], \nonumber\\
E_s=\frac{n}{2}\left[C_\rho+\frac{\pi^2}{2 k_F
E}-\frac{C_\delta^{eff}} {1+C_\delta^{eff}H^{(2)}}
\left(\frac{M}{E}\right)^2 \nonumber \right],
\end{eqnarray}
which agree with results obtained in relativistic field models
with structureless nucleons, with exception of the couplings
$C_\phi^{eff}=\left(g_\phi Q/m_\phi\right)^2$ ($\phi=\sigma,
\delta$) which includes the factor $Q$ defined below Eq.
(\ref{LPARAM1}). Taking into account that $Q$ depends on the
medium properties through $m^\ast_q$ and $R$, density dependent
effective couplings have been obtained, due to the quark structure
of nucleons proposed in the model.

\begin{figure}[ht]
\centering
\includegraphics[width=0.8\textwidth]{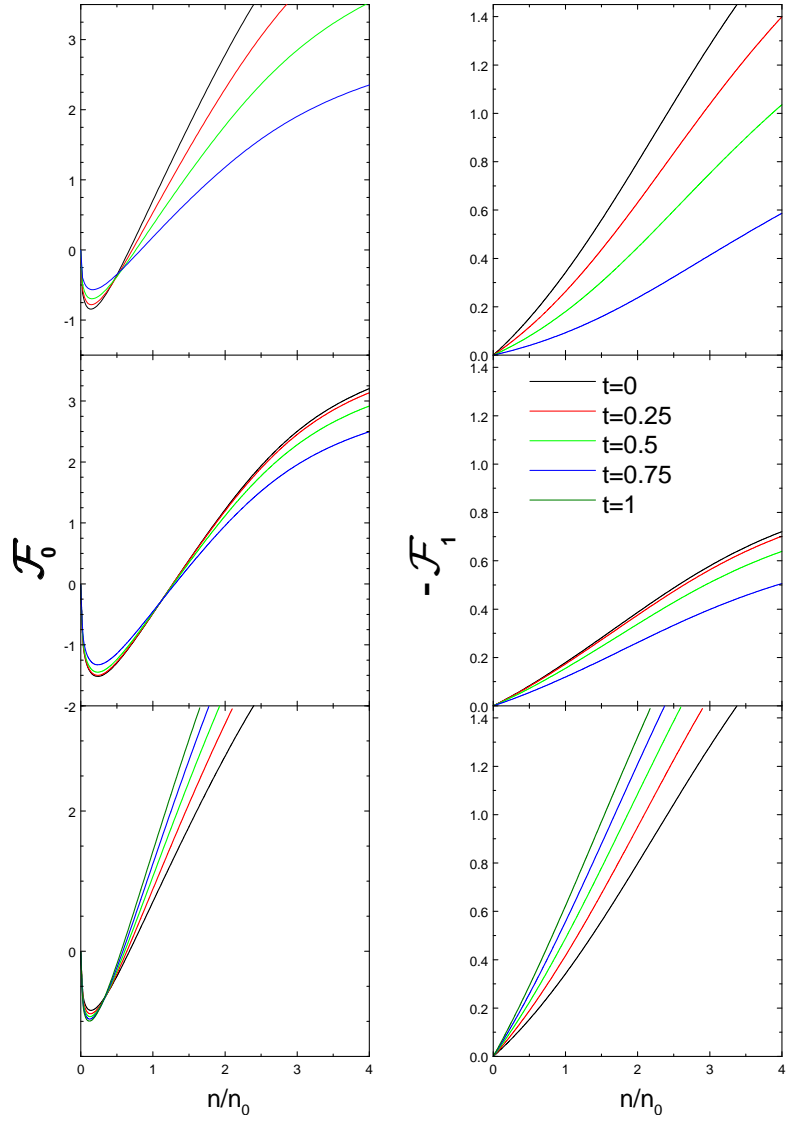}
\caption{\footnotesize{The adimensional Landau parameters for the
CC case. In the left (right) panels the scalar (vector) components
for the $p\,p$, $p\,n$, and $n\,n$ pairs are respectively plotted
from top to bottom, at several fixed values of $t$. The line
convention valid for all the cases is indicated in the central
right panel, it must be noted that $t=1$ is relevant only for the
$n\,n$ amplitude.}}\label{FIG3}
\end{figure}

\begin{figure}[ht]
\centering
\includegraphics[width=0.8\textwidth]{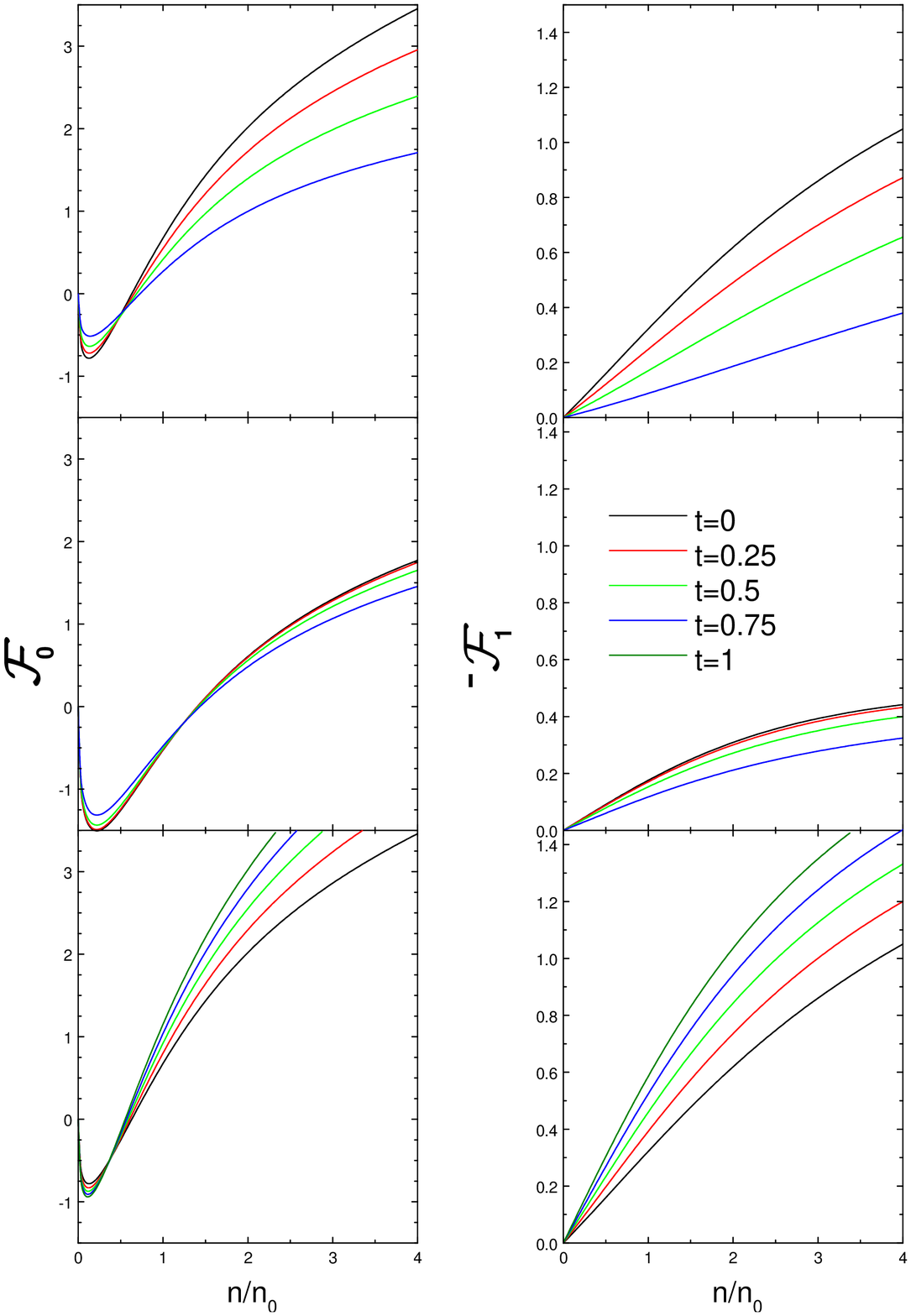}
\caption{\footnotesize{Idem as Fig.\ref{FIG3}, but for the NC
case. }}\label{FIG4}
\end{figure}

The density dependence of the adimensional Landau parameters are
plotted in Figs. \ref{FIG3} (CC) and \ref{FIG4} (NC). At
sufficient low densities  all scalar parameters
$\mathcal{F}_{p\,p}^{0}$, $\mathcal{F}_{n\,p}^{0}$, and
$\mathcal{F}_{n\,n}^{0}$  are negative, reflecting the attractive
character of the effective nucleon interaction. Therefore
instabilities in the equation of state can appear in this density
range. Comparing Figs. \ref{FIG3} and \ref{FIG4} we appreciate
that, for densities $n/n_0\lesssim 1$, the general trend of
excluded volume correlations is to slightly increase the
nucleon-nucleon attraction. On the other hand in the range $n/n_0
> 1$ the  volume corrections  enhance the repulsion among nucleons
as compared with the respective NC cases. This fact is expected
since the available volume per nucleon decreases, resulting in a
non negligible compression of the bags at higher densities
\cite{OUR}.\\
The finite volume effects  are more evident for the
$\mathcal{F}_{b\,b'}^{1}$ ($b, b'=p, n$) components, enhancing
their absolute values for all the range of
densities, specially for  $n/n_0 > 1$.\\
Because the adimensional Landau parameters contain density
dependent factors, $\mathcal{F}_{p\,p}^{0,\,1}$ and
$\mathcal{F}_{p\,n}^{0,\,1}$ decrease for growing isospin
asymmetry $t$ as they are a measure of the strength of the
in-medium proton interaction.
The opposite is true for $\mathcal{F}_{n\,n}^{0,\,1}$.\\
The low density limit of CC results qualitatively agree with
others calculations \cite{MATSUI,CAILLON},
both for nuclear symmetric matter and for the pure neutron case.\\
It can be seen that in the sub-nuclear realm of the scalar
interaction, the in-medium $p-n$ strength overrides the $p-p$ and
$n-n$ components. Furthermore the $n-p$ attractive effect has a
wider density range, extending beyond the normal value $n_0$.

\begin{figure}[ht]
\centering
\includegraphics[width=0.8\textwidth]{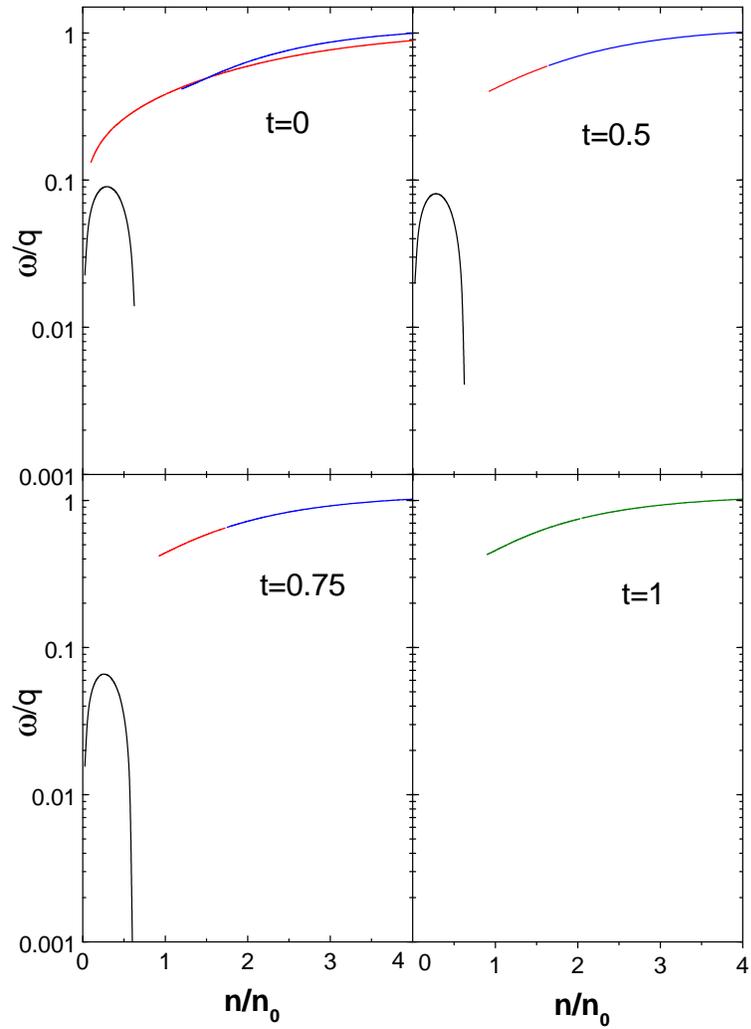}
\caption{\footnotesize{Dispersion relation for collective modes in
the CC case, for several values of the asymmetry parameter $t$.
The low density solution represents the unstable isoscalar mode
(black line). The remaining stable modes are depicted with red or
blue lines for isovector or isoscalar character, respectively.
}}\label{FIG5}
\end{figure}

\begin{figure}[ht]
\centering
\includegraphics[width=0.8\textwidth]{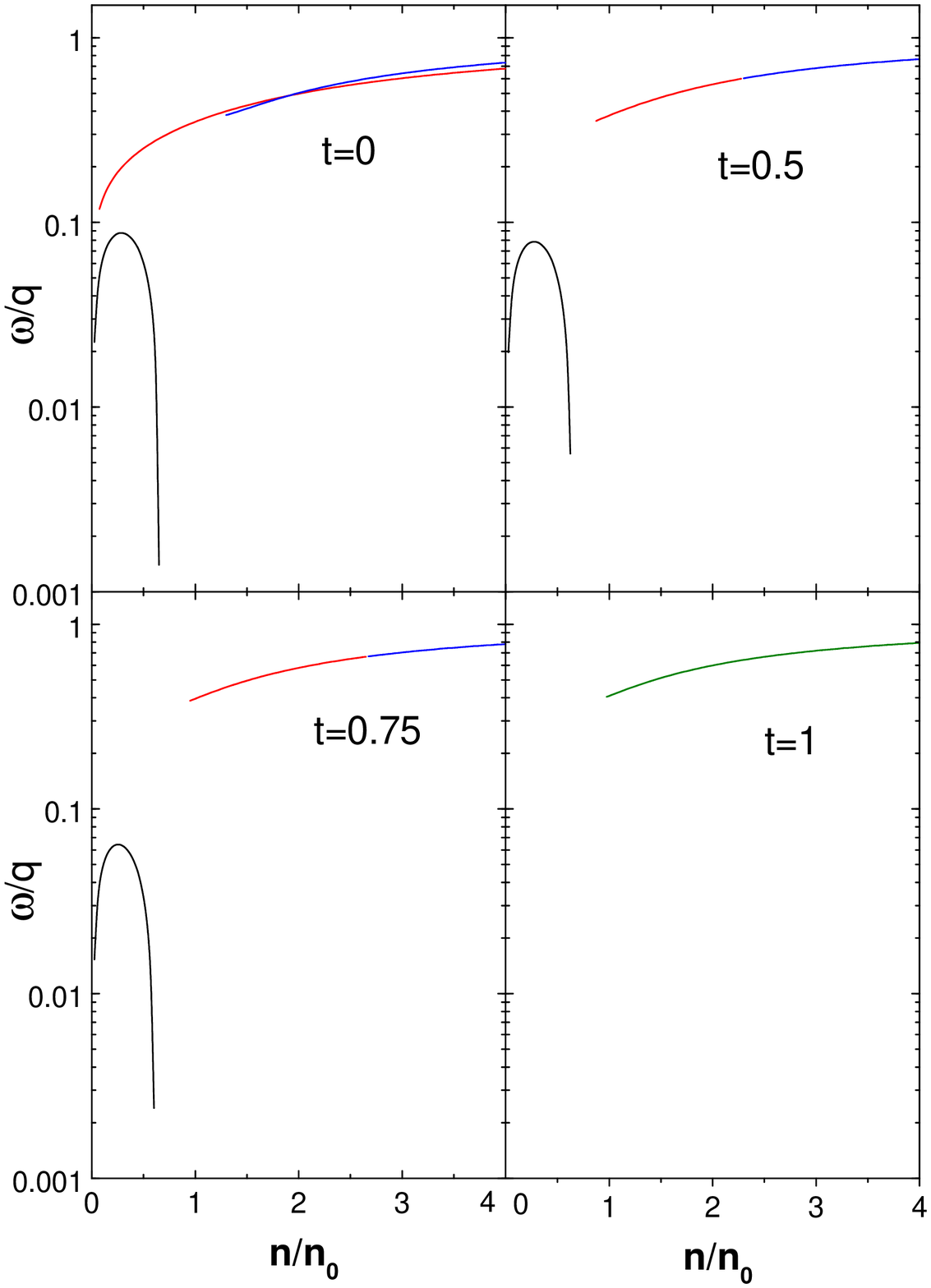}
\caption{\footnotesize{Idem as Fig.\ref{FIG5}, but for the NC
case.}}\label{FIG6}
\end{figure}

Collective quantum fluctuations give rise to proper modes which
are solutions of the eigenvalue equation (\ref{EIGEN}). The
corresponding dispersion relations $(\omega/q)$ are displayed in
Figs. \ref{FIG5} (CC) and \ref{FIG6} (NC), for some typical values
of the isospin asymmetry $t$. It must be pointed out that stable
modes satisfy $s_{p,n}> 1$, whereas pure imaginary values
correspond to unstable propagation. In the last case the quantity
$|\omega/q|$ has been plotted. With exception of nearly pure
neutron matter, unstable modes are always present at very low
densities and therefore they are practically insensitive to finite
volume effects.

As it was mentioned earlier, collective modes can be classified as
isoscalar and isovector, according to proton and neutron
vibrations being in phase or in opposition, respectively. \\
Unstable modes are found to be isoscalar in character, reflecting
the fact that both isospin components simultaneously undergo a
liquid-vapor phase transition, leading to cluster formation
\cite{BARAN}. This associated mechanical instability is evidenced
by the negative sign of the nuclear compressibility in this
density domain. Since $\kappa$ comprises density variations at
fixed isospin asymmetry
$t$, Eq.(\ref{KAPPA}), it preserves the proton to neutron ratio.\\
It is easily verified that for the symmetric case isoscalar modes
(stable or unstable) also conserve this ratio, namely $\varrho$
equals ${(\delta n^p/\delta n^n)}_t =(n^p/n^n)$ for $t=0$.
This is no longer true for iso-asymmetric matter.\\
For growing asymmetry $t$ these unstable modes show a decreasing
amplitude, because the repulsive
n-n interaction progressively dominates, until the instability vanishes.\\
For stable eigensolutions $(\omega/q)$ represents the zero sound
velocity $V_s$. In the iso-symmetric case there are two stable
modes, both in CC and NC instances. The isovector branch appears
at very low densities, whereas the isoscalar one starts at $n/n_0
> 1$ . For the CC plot they cross each other at $n/n_0 \simeq
1.51$ ($n/n_0 \simeq 1.9$ for NC), but keep their own character
because isoscalar and mechanical oscillations preserve the same
proton to neutron ratio for $t=0$. Thus, they do not couple to
isovector fluctuations, which are related to species
separation \cite{BARAN}.\\
In general for iso-asymmetric matter there is only one stable
branch which has a mixed character. In fact, the ratio of proton
to neutron amplitudes $\varrho$ changes smoothly from negative
(isovector) to positive (isoscalar) as the density increases. When
the nucleon finite size is considered (CC) the change of
character takes place at lower densities than in the point-like case (NC),
as can be appreciated from figures \ref{FIG5} and \ref{FIG6}, respectively.\\
As mentioned before, in asymmetric matter the ratio $\varrho$ for
isoscalar modes (either stable or not) do not follow the
constraint of keeping $t$ constant, as mechanical oscillations do
\cite{BARAN}. In fact, the asymmetry of the medium induces a more
complicated scenario where some chemical component is also
involved, i.e. the relative proton to neutron concentration is
modified along the quantum mode. The mixed iso-character of the
stable modes at finite $t$ can be ascribed to this cause. At low
densities these collective modes are isovector like, indicating a
significative species separation. As the density grows, pressure
(mechanical) effects become dominant and induce the change to
isoscalar like character.\\
For pure neutron matter only one stable mode has been found in
this range of densities.\\
Concerning the velocity of propagation, a rough $30 \%$ increment
in the CC results for $V_s$ are observed in the high density
domain as compared to the NC ones. As an artifact of the
approximation used to solve the Landau's kinetic equation, it is
found that $V_s$ approaches the velocity of light at densities
$n/n_0 \gtrsim 4$, in the CC approach. However, for such high
densities the dissipative effects are non-negligible, resulting in
a more involved dynamics which is beyond the scope of the present
work.

\section{Conclusions}

We have studied the thermodynamical properties and collective
modes of cold asymmetric nuclear matter. This has been performed
within a model of structured nucleons, taking into account an
appropriate normalization of the nucleon fields to prevent
overlapping of the quark confining regions at high densities
\cite{OUR}. Many body properties of hadronic matter are
interpreted in a Fermi liquid picture with quasiparticles and
collective modes.\\
Within this framework the relativistic Landau parameters for
iso-asymmetric nuclear matter have been evaluated, and an explicit
relation with the nuclear compressibility $\kappa$ and symmetry
energy $E_s$ has been obtained. The overall trend of excluded
volume correlations at densities $n>n_0$ is to develop an stiffer
equation of state, relative to the case where they are absent.\\
Different quantities such as density dependence of the symmetry
energy, its slope, and the asymmetry compressibility $K_{assy}$
have been evaluated, obtaining qualitative agreement with
empirical
estimates. \\
It is argued that short range hadronic correlations in the dense
medium preceding the deconfinement phase transition, which has
been parametrized in the present work as excluded volume
corrections, contribute to understand the high
neutron star mass measured recently.\\
We have also applied the present formalism to the Landau's
collisionless kinetic equation for small fluctuations around the
Fermi level. We have obtained the eigenvalue equation for
instability and zero sound modes within this scheme, and have
discussed the propagation of these modes in the dense hadronic
medium.\\
The behavior of collective excitations can have important
consequences in heavy ion reactions, where the formation of
fragments is chiefly governed by isoscalar fluctuations.

\newpage
\begin{table}[ht]
\centering
\begin{tabular}{|l|c|c|c|c|} \hline
Case &  $g_{\sigma}^{u,d}$ &
$g_{\omega}^{u,d}$ & $g_{\delta}^{u,d}$ & $g_{\rho}^{u,d}$ \\ \hline
CC&5.76314&2.78280&5.75150&4.3500\\
NC&5.99339&3.00770&5.42075&4.5000\\ \hline
\end{tabular}
 \caption{\footnotesize{The quark-meson couplings
 used in the case with excluded volume correction (CC) and without it
(NC).}}\label{TABLEI}
\end{table}

\newpage

\end{document}